\documentclass[12pt]{article}
\usepackage{amsmath,amsbsy,amssymb,graphics}
\usepackage{graphicx,epsfig}
\def\beq{\begin{equation}}
\def\eeq{\end{equation}}
\def\be{\begin{equation}}
\def\ee{\end{equation}}
\def\bea{\begin{eqnarray}}
\def\eea{\end{eqnarray}}
\def\nnb{\nonumber}

\newcommand{\gsim}{\lower.7ex\hbox{$\;\stackrel{\textstyle>}{\sim}\;$}}
\newcommand{\lsim}{\lower.7ex\hbox{$\;\stackrel{\textstyle<}{\sim}\;$}}

\begin{document}

\begin{center}
 \vspace{0.2cm}
 {\large \bf Signal of New Physics and Chemical Composition
 of Matter in Core Crossing Neutrinos }

\vspace{0.6cm} {\large \bf Wei Liao }

\vspace{0.3cm} { Institute of Modern Physics, P.O. Box 532\\
 East China University of Science and Technology \\
 130 Meilong Road, Shanghai 200237, P.R. China\\

 \vskip 0.3cm

 Center for High Energy Physics\\
 Peking University, Beijing 100871, P. R. China }

\end{center}
\begin{abstract}
 \vskip 0.2cm
 We consider non-standard matter effect in flavor conversion of
 neutrinos crossing the core of the Earth. We show that oscillation
 of core crossing neutrinos with $E \gsim 0.5$ GeV can be well
 described by a first order perturbation theory. We show that
 due to non-standard matter effect varying chemical composition
 in the Earth can modify the neutrino flavor conversion by $100\%$.
 Effects of CP violating phases in non-standard Neutral Current
 interactions are emphasized in particular.

\end{abstract}


\section{Introduction}\label{sec1}

 It is well known that non-standard interaction(NSI)
 can induce non-standard matter effect for neutrino oscillation in medium.
 Neutrino flavor conversion induced by non-standard matter effect was
 proposed as a candidate solution to the solar neutrino anomaly~\cite{msw1}.
 The present experiments told us that LMA MSW solution~\cite{msw1,msw2} with the
 standard interaction is the solution to the solar neutrino
 problem~\cite{sno1,sno2,sk,kamland}. Non-standard matter effect is small
 in oscillation of solar neutrinos. However
 non-standard matter effect can be much larger for neutrinos
 with high energy ($E \gsim 10$ GeV), e.g. for long baseline neutrinos,
 atmospheric neutrinos, cosmic neutrinos from the galactic or extra-galactic sources.
 This is because flavor conversion induced by flavor mixing in vacuum
 decreases as energy increases while the matter effect does not decrease with energy.
 Previous works on effect of NSI in neutrino oscillation include
~\cite{some2,NSI-longbase}.

 Non-standard matter effect can be induced by non-standard
 Neutral Current interaction of neutrinos with electron, proton and neutron.
 In this respect non-standard matter effect in neutrino oscillation
 is not only a way to probe physics beyond the Standard Model but also
 a way to probe chemical composition in matter. Incorporating
 non-standard matter effect in neutrino oscillation introduces
 more CP violating phases in the Hamiltonian. These CP violating
 phases interfere with the CP violating phase in vacuum
 and can give interesting phenomena. In matter with varying chemical
 composition these CP violating phases can contribute with different combinations
 in observables.

 It is the purpose of the
 present article to study the effect of varying chemical composition
 in the Earth. Effects of CP violating phases in the
 non-standard interaction will be analyzed in particular.
 In section \ref{sec2} we show that oscillation of core crossing neutrinos
 in the Earth can be well described by a first order perturbation theory which was
 developed in a previous paper by the author. Scenarios with different CP violating phases
 and varying composition in the Earth are shown. In section \ref{sec3} we show the
 effect of non-standard interactions and CP violating phases in the non-standard
 interactions. We summarize and comment in section \ref{sec4}.
 We do analysis using the density profile
 of the Preliminary Earth Model(PREM) ~\cite{PREM}.

 \section{Non-standard matter effect in the Earth} \label{sec2}
 We consider oscillation of three flavors of neutrinos:
 $\psi=(\nu_e, \nu_\mu,\nu_\tau)$.  The evolution equation is
  \bea
  i \frac{d}{d x} \psi(x) = H(x) \psi(x), \label{evol1}
 \eea
 where
 \bea
  &H(x) = H_0 + V(x),
 \label{evol1a}\\
 & H_0 =\frac{1}{2 E} U ~ \textrm{diag}\{0, \Delta m^2_{21}, \Delta m^2_{31}\}~
 U^\dagger.
  \label{evol1b}
 \eea
 $V(x)$, a $3\times 3$ matrix, is the potential term
 accounting for the matter effect. $U$ is the $3\times 3$
 neutrino mixing matrix in vacuum. $U$ is parameterized using
 standard parameters $\theta_{12}$, $\theta_{13}$, $\theta_{23}$ and
 $\delta_{13}$, the CP violating phase.

  In the presence of non-standard NC interaction
  the potential term can be written as follows
 \bea
 V(x)= \textrm{diag}\{V_e,0,0 \} +\begin{pmatrix}0 & V_{e\mu} & V_{e\tau} \cr
        V_{\mu e} & V_{\mu\mu} & V_{\mu \tau} \cr
        V_{\tau e} & V_{\tau \mu} & V_{\tau \tau}\end{pmatrix},
 \label{NonSHamil}
 \eea
 where $V_e=\sqrt{2} G_F N_e$ is the potential with standard charged current
 interaction. $G_F$ is Fermi constant and $N_e$ is electron
 number density.
 $V_{kl}$ is from non-standard NC interaction. $V_{lk}^*=V_{kl}$ because the Hamiltonian
 is hermitian.  $x$ dependence in $V_{kl}$ has been suppressed in Eq.
 (\ref{NonSHamil}). $V_{ee}$ has been made zero in our convention. This is
 achieved by shifting the phases of neutrinos:
  $\nu_l \to e^{-i \int dx ~V_{ee} } ~\nu_l$.

  In this convention $V_{kl}$ is
 \bea
 V_{kl} &&=\sqrt{2} G_F ~ \sum_{s=e,p,n} (f^s_{kl}-f^s_{ee}) N_s \nnb \\
  &&= V_e ~[\sum_{s=e,p,n} (f^s_{kl}-f^s_{ee})
   +(f^n_{kl}-f^n_{ee}) R_n], \label{NonSHamilA},
 \eea
 where
 \bea
 R_n=(N_n-N_e)/N_e. \label{Ratio}
 \eea
 $f^e_{kl}$,~$f^p_{kl}$ and $f^n_{kl}$ are the dimensionless strengths of
 non-standard four Fermion interactions
   $\sqrt{2} ~f^s_{kl} ~G_F ~{\bar s} \gamma_\mu s ~{\bar \nu_k} \gamma^\mu
 \nu_l$. $N_p$ and $N_n$ are number densities of proton and neutron in
 matter. In obtaining the second line of Eq. (\ref{NonSHamilA}) $N_e=N_p$ in neutral
 matter has been used.

   We can re-write $V(x)$ as
  \bea
  V(x)=V_e(x) \begin{pmatrix} 1 & \epsilon_{e\mu} & \epsilon_{e\tau} \cr
   \epsilon_{\mu e} & \epsilon_{\mu \mu} & \epsilon_{\mu \tau} \cr
   \epsilon_{\tau e} & \epsilon_{\tau \mu} & \epsilon_{\tau \tau}
   \end{pmatrix}. \label{NonSHamilB}
  \eea
 where $\epsilon_{kl}=V_{kl}/V_e$. We can write
 \bea
 \epsilon_{kl}=\epsilon^0_{kl} (1+ R_n ~r_{kl} ~e^{-i
 \phi_{kl}} ), \label{NonSHamilC}
 \eea
 where
 \bea
 \epsilon^0_{kl} =\sum_{s=e,p,n} (f^s_{kl}-f^s_{ee}),~
 r_{kl} ~e^{-i \phi_{kl}}= (f^n_{kl}-f^n_{ee})/\epsilon^0_{kl}.
 \label{NonSHamilD}
 \eea
 $\epsilon^0_{kl}$ is constant in matter.
 $\epsilon_{kl}$ depends on the chemical composition in matter and
 may have $x$ dependence in neutrino trajectory. $r_{kl}$ and
 $\phi_{kl}$ are real numbers. $\epsilon^0_{ee}=0$ and $r_{ee}=0$ in
 our convention. $\epsilon^0_{kl}=\epsilon^{0*}_{lk}$ and
 $\phi_{kl}=-\phi_{lk}$ because of the hermiticity of $V$.

 Constraints on $\epsilon_{kl}$ come from direct test on NSI~\cite{NuTeV,pdg} and the
 neutrino oscillation experiments. Test on NSI
 can not be directly translated to constraint on $\epsilon_{kl}$.
 These constraints have been discussed in our previous work
 \cite{liao}. It was shown that present constraints are
 $|\epsilon_{\mu \mu}|, |\epsilon_{\tau \tau}| \lsim 10^{-2}$ and
$|\epsilon_{\mu e}|, |\epsilon_{\mu \tau}| \lsim 10^{-2}$,
 $|\epsilon_{e \tau}| \lsim 10^{-1}$~\cite{liao,constraint,constraintA}.

 It is clear that $\epsilon^0_{kl}$ introduces three CP
 violating phases in addition to the phase $\delta_{13}$ in matrix $U$.
 They are phases of $\epsilon^0_{e\mu,e\tau, \mu \tau}$.
 Furthermore $\phi_{e \mu, e \tau,\mu \tau}$ become independent phases
 in case that chemical composition varies in matter. So in matter with varying
 chemical composition, i.e. $R_n$ not a constant, we have seven
 physical CP violating phases in total. They can give interesting phenomena
 in neutrino oscillation. In Earth matter $R_n$ is estimated\cite{ls}
 \bea
  R_n =\left \{
 \begin{matrix} 0.024, & \textrm{~~~mantle} \cr
          0.146, & \textrm{core}
 \end{matrix}    \right. \label{RatioA}
 \eea
 We will use numbers in (\ref{RatioA}) in our analysis in
 the present article. We will concentrate on neutrinos with
 core crossing trajectories.

 It was shown in a previous work that oscillation
 of neutrinos in the Earth can be well described by a
 first order perturbation theory~\cite{liao,comment}.
 The theory was analyzed with the
 assumption that $\epsilon_{kl}$ is a constant in
 neutrino trajectory. We show in this section that
 this theory works perfectly well taking into account the
 fact that chemical composition in the core and in the mantle
 are different.

   \begin{figure}
\begin{flushleft}
\includegraphics[height=6.cm,width=8cm]{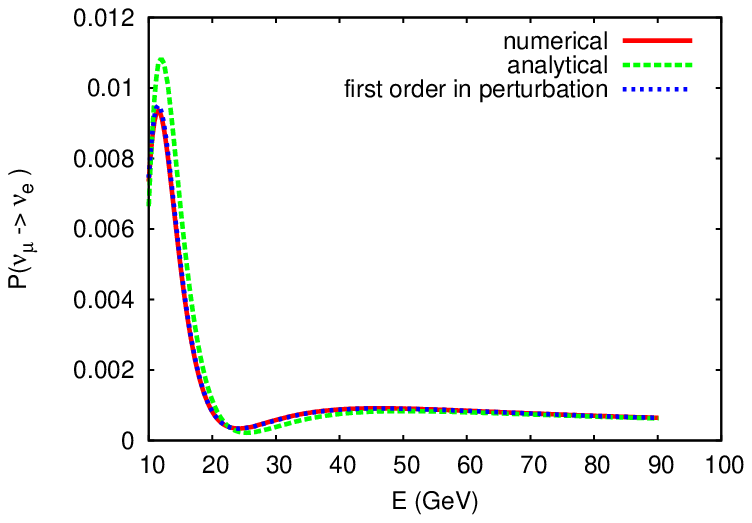}
\end{flushleft}
\begin{flushright}
\vskip -6.5cm
\includegraphics[height=6.cm,width=8cm]{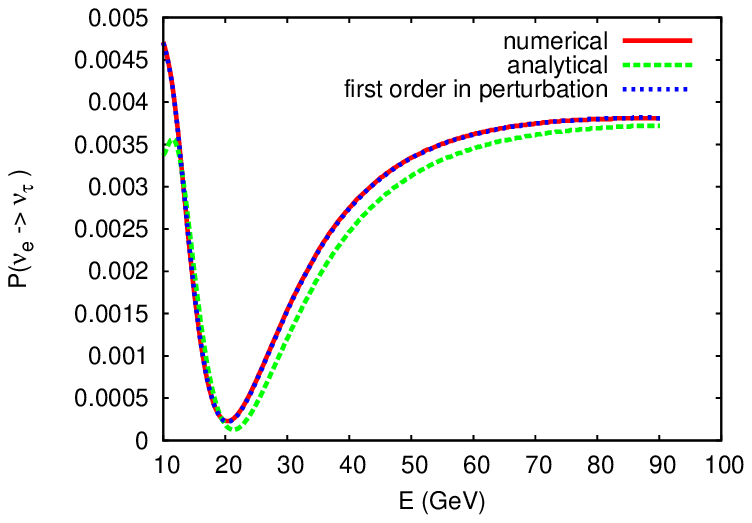}
\end{flushright}
\vskip 0.0cm \caption{\small Left $P(\nu_\mu \to \nu_e)$ versus
 energy; right $P(\nu_e \to \nu_{\tau})$ versus energy.
 $L=12000$ km, $\Delta m^2_{21}=8. \times 10^{-5}$ eV$^2$,
 $\Delta m^2_{32}= 3. \times 10^{-3}$ eV$^2$. $\sin^2 2\theta_{23}=1$,
 $\tan^2 \theta_{12}=0.41$, $\sin^2 2\theta_{13}=0.01$, $\delta_{13}=\pi/6$.
 $\epsilon^0_{e\mu}=0.01 ~e^{-i \pi/20}$,
  $\epsilon^0_{e\tau}=0.04 ~e^{-i \pi/3}$, $\epsilon^0_{\mu\tau}=0.01 ~e^{-i
  \pi/20}$. $r=5$, $\phi_{e\mu}=\phi_{\mu \tau}=\pi/2$, $\phi_{e\tau}=0$.
 PREM density profile is used for computation in this figure and
 all remaining figures in this article.}
 \label{PVsE}
 \end{figure}

     \begin{figure}
\begin{flushleft}
\includegraphics[height=6.cm,width=8cm]{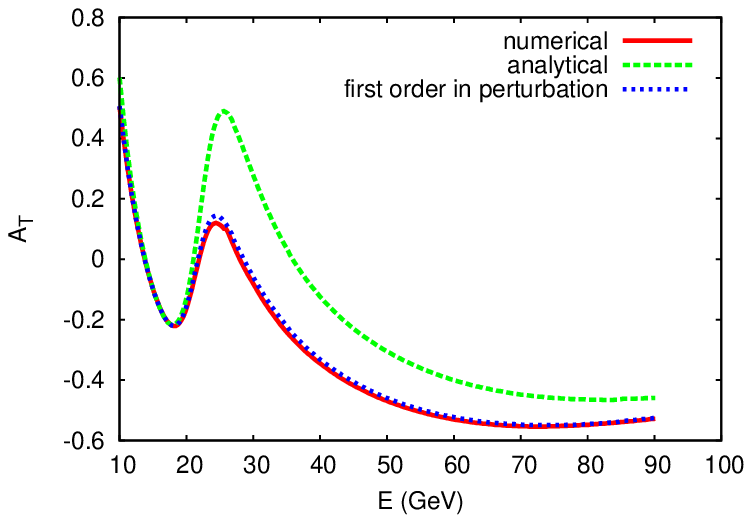}
\end{flushleft}
\begin{flushright}
\vskip -6.5cm
\includegraphics[height=6.cm,width=8cm]{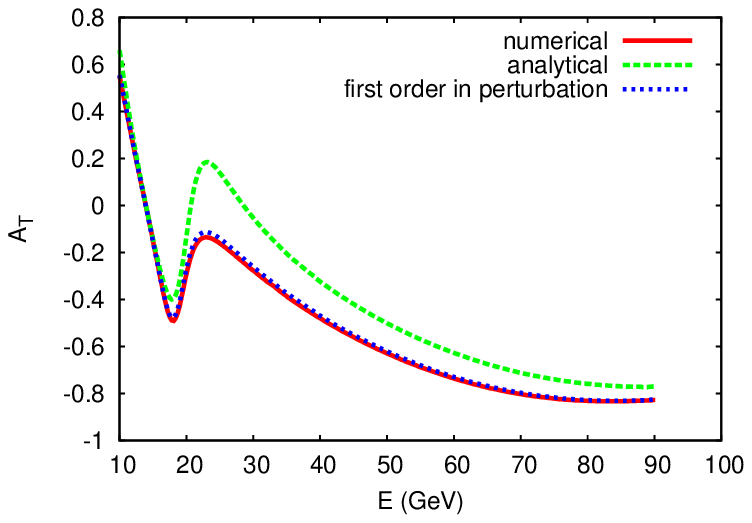}
\end{flushright}
\vskip 0.0cm \caption{\small Time reversal asymmetry, $A_T$, versus
 energy. Left $r=5$; right $r=2$. Other parameters
 are the same as in Fig. \ref{PVsE}.
 }
 \label{TsVsE}
 \end{figure}

 We quickly review the perturbation theory.
 We denote $L$ as the length of neutrino trajectory in the Earth.
 For core crossing neutrinos($L \gsim 10690$ km)
 we write the evolution matrix $M$ as
    \bea
   M=M_3 M_2 M_1, \label{CoreMantle}
   \eea
 where $M_2$ is the evolution matrix in the core and $M_{1,3}$
 are evolution matrices in the mantle.  $0<x <L_1$ and
 $L_2 < x < L$ are the parts of trajectory in the mantle;
 $L_1 <x <L_2$ is the part of trajectory in the core.
 We average potential in the mantle and in the core
 separately
 \bea
 {\bar V}_i= \frac{1}{L_i-L_{i-1}} \int^{L_i}_{L_{i-1}} ~dx ~V(x),
 ~~~i=1,2,3,
 \label{avepotenb}
 \eea
 where $L_3=L$.
 Using ${\bar V}_i$ we define the average Hamiltonian
 \bea
 {\bar H}_i= H_0 + {\bar V}_i, ~~i=1,2,3 \label{defH}
 \eea
 and get eigenvector and mixing matrix
 \bea
 {\bar H} U_{mi} = U_{mi} ~\frac{\Delta_i}{2 E}, ~~i=1,2,3. \label{defU}
 \eea
 $\Delta_i$ is a vector. The evolution matrix $M_i$ is expressed as
 \bea
 M_i &&=U_{mi} ~e^{- i \frac{\Delta_i}{2 E} (L_i-L_{i-1})}(1- i C^i) ~U^\dagger_{mi},
 ~~i=1,2,3 \label{evol1}
 \eea
 $C^i$ is a $3\times 3$ matrix accounting for the non-adiabatic
 transition:
 \bea
 C^i=\int^{L_i}_{L_{i-1}} dx ~e^{i \frac{\Delta_i}{2 E} x} ~U^\dagger_{mi} ~\delta V_i(x)
 ~U_{mi}  ~e^{-i \frac{\Delta_i}{2 E} x} \label{evol2},
 \eea
 where
 \bea
 \delta V_i(x)= V(x)- {\bar V}_i.
 \eea
 It is clear that $(C^i)^\dagger =C^i$ holds.

 In \cite{liao} we have discussed in detail that this theory is indeed doing expansion
 using small quantities.  $C^i_{jk}(j\neq k)$ is suppressed by small
 quantities for neutrinos with $E \gsim 0.5 $ GeV. Second order
 effect is of order ${\cal O}(C^2)$ and is further suppressed.

 In Fig. \ref{PVsE} we compare the result of numerical computation
 with that computed in the first order perturbation theory,
 i.e. using Eqs. (\ref{CoreMantle}) and (\ref{evol1}).
 For simplicity we have set
 \bea
 \epsilon_{\mu \mu}=\epsilon_{\tau \tau}=0,~~
 r_{e\mu}=r_{e\tau}=r_{\mu \tau}= r.
 \eea
 We see that result computed using the perturbation theory is
 in remarkable agreement with the numerical result.

 We also show the zeroth order result, i.e. result computed by
 setting $C_i$ zero in Eq. (\ref{CoreMantle}).
 The zeroth order result is an analytical result
 computed using average potentials in the core
 and in the mantle separately. We see that the analytical result
 is not a bad approximation to the oscillation pattern. It
 qualitatively describes the neutrino oscillation pattern and
 can help a lot when making qualitative discussions.

 In Fig \ref{TsVsE} we show plot of time reversal asymmetry versus
 energy. $A_T$ is defined as
 \bea
  A_T &&= \frac{P(\nu_e \to \nu_\mu)-P( \nu_\mu \to \nu_e)}
 {P(\nu_e \to \nu_\mu)+P(\nu_\mu \to \nu_e)}. \label{AT}
 \eea
 Again we see that the first order perturbation theory gives a
 perfect description of the oscillation pattern. The analytical
 result gives a qualitatively good approximation to neutrino
 oscillation. It can help in making qualitative discussions.

\section{Flavor conversion of core crossing neutrinos}\label{sec3}

 In this section we illustrate the effect of CP violating phases of
 NSI in neutrino oscillation.

 As shown in the last section,
 oscillation of core-crossing neutrino can be qualitatively
 described by approximation which uses average densities in the
 mantle and in the core separately. This is an analytical
 description. We use this description to simplify the discussion
 and see the effect of $\phi_{kl}$ in neutrino oscillation.

 It is easier to discuss in the large energy region where we
 can re-write the Hamiltonian as
 \bea
 H= V_0 + H_1, \label{HamilA}
 \eea
 where
 \bea
 V_0=\textrm{diag}\{V_e,0,0 \} ,~~ H_1=V-V_0+H_0. \label{HamilB}
 \eea
$V_0$ is taken as the leading term in the Hamiltonian. $H_1$ is
taken as perturbation. $V-V_0$ is for the non-standard matter effect
and $H_0$ is the Hamiltonian in vacuum. $H_0$ decreases as energy
increases.

     \begin{figure}
\begin{flushleft}
\includegraphics[height=6.cm,width=8cm]{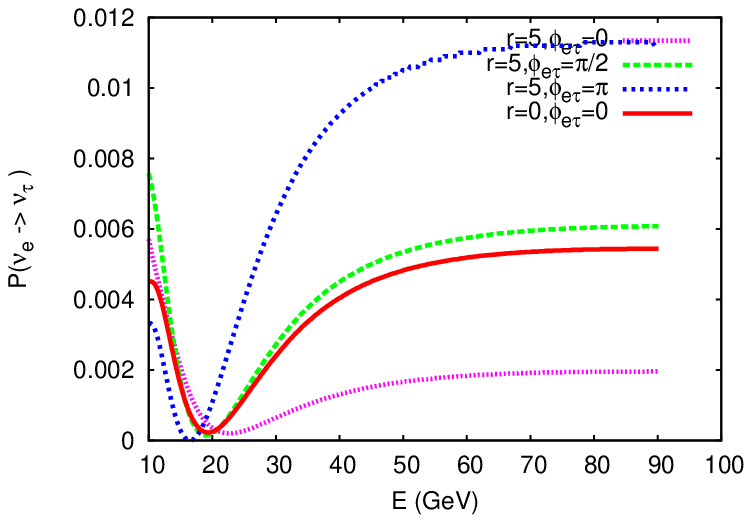}
\end{flushleft}
\begin{flushright}
\vskip -6.5cm
\includegraphics[height=6.cm,width=8cm]{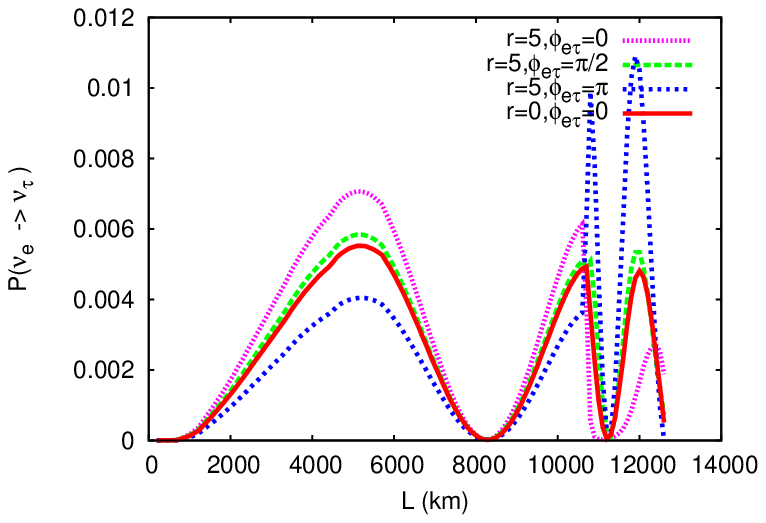}
\end{flushright}
\vskip 0.0cm \caption{ \small $P(\nu_e \to \nu_\tau)$ versus
 energy, $L=12000$ km; $P(\nu_e \to \nu_\tau)$ versus
 distance L, $E=50$ GeV. $\phi_{e\mu}=\phi_{\mu \tau}=\pi/2$. Other parameters
 are the same as in Fig. \ref{PVsE}.
 }
 \label{PVsEB}
 \end{figure}

 Using the average potentials in the core and in the mantle
 we can get the evolution matrix using perturbation in $H_1$.
 As an example, $\nu_e \to \nu_\tau$ amplitude is
 \bea
 A(\nu_e \to \nu_\tau) && \approx \frac{(H^m_1)_{\tau e} }{V^m_e}(e^{-i \varphi_1} -1)
 + \frac{(H^c_1)_{\tau e} }{ V^c_e} ( e^{-i \varphi_c }
 -1) e^{-i \varphi_1} \nnb \\
 && + \frac{(H^m_1)_{\tau e} } {V^m_e} ( e^{-i \varphi_1} -1)
 e^{-i (\varphi_c + \varphi_1) }, \label{evol2b}
 \eea
 where $\varphi_c= V^c_e (L_2- L_1)$ and $\varphi_1= V^m_e L_1$.
 $H^m_1$ and $H^c_1$ are averages of $H_1$ in the mantle and in the core.
 $V^m_e$ and $V^c_e$ are averages of $V_e$ in the mantle and in the core.
 The property of approximately symmetric density profile in the Earth
 has been used in Eq. (\ref{evol2b}).
 It can be written as
 \bea
 A(\nu_e \to \nu_\tau) \approx \frac{(H^m_1)_{\tau e} }{V^m_e}(
 e^{-i \varphi} -1)+( \frac{H^c_1 }{ V^c_e}- \frac{H^m_1 }{V^m_e} )_{\tau e}
 ( e^{-i \varphi_c } -1) e^{-i \varphi_1},
 \label{evol2c}
 \eea
 where $\varphi=\varphi_c+2 \varphi_1$.

 Neglecting terms of order ${\cal O} ( \frac{\Delta m^2_{21} } {2 E V_e} ,
 \frac{\Delta m^2_{31} }{2 E V_e} \sin\theta_{13} )$, we get
 \bea
  A(\nu_e \to \nu_\tau) && \approx \epsilon^0_{\tau e} ~(1+ 0.024 ~r_{\tau e}
  ~e^{-i \phi_{\tau e}}) (e^{-i \varphi} -1) \nnb \\
  && + 0.122 ~\epsilon^0_{\tau e}
  ~r_{\tau e} ~e^{-i \phi_{\tau e}}(e^{-i \varphi_c} -1) e^{-i \varphi_1}.
 \label{evol2d}
 \eea
 Eq. (\ref{RatioA}) has been used in obtaining Eq. (\ref{evol2d}).
 $A(\nu_e \to \nu_\tau)$ is determined by
 $\epsilon^0_{\tau e}$ modulated by contribution of $r_{\tau e}$ and
 $\phi_{\tau e}$. For neutrinos which do not cross the core of the
 Earth the second term in the
 r.h.s. of Eq. (\ref{evol2d}) is absent.

 One can see clearly in
 Eq. (\ref{evol2d}) that if $r_{\tau e}=0$ the transition amplitude,
 $A(\nu_e \to \nu_\tau)$, is determined by $\epsilon^0_{\tau e}$ and modulated by factor
 $e^{-i \varphi} -1$. Hence $P_{e\tau}$ is proportional
 to function $\sin^2(\varphi/2)$. In the right panel of Fig.
 \ref{PVsEB} we see plot for this case. For $r_{\tau e}=0$ and
 $\phi_{\tau e}=0$, $P_{e\tau}$ has three peaks with roughly equal heights, as
 expected. For $r_{\tau e}=5$, $P_{e\tau}$ is considerably changed.
 When $\varphi_{\tau e}=0$ and $\varphi_{\tau e}=\pi$,
 $P_{e\tau}$ is considerably modified
 around the third peak (for core-crossing neutrinos).
 The height in this peak is quite different
 from that in the first peak (for neutrinos crossing the mantle only).
 This is quite different from the case with $r_{\tau e}=0$.
 In the left panel of Fig. \ref{PVsEB} we show the plots of
 transition probability versus energy. Again we see the effect of
 $r_{\tau e}$ and $\varphi_{\tau e}$. $P_{e\tau}$ is also slightly modified
 in the first peak when $r_{\tau e} \neq 0$. This is because
 of the correction by $r_{\tau e}$ to the Hamiltonian in the mantle,
 as shown in the first term in the r.h.s of Eq. (\ref{evol2d}).
 Comparing with the transition
 probability in neutrinos crossing the mantle, the core-crossing
 neutrino events encode the information of $r_{\tau e}$ and
 $\varphi_{\tau e}$. And effect of $\epsilon^0_{kl}$ and $r_{kl}$
 are distinctly different in oscillation probability.

     \begin{figure}
\begin{flushleft}
\includegraphics[height=6.cm,width=8cm]{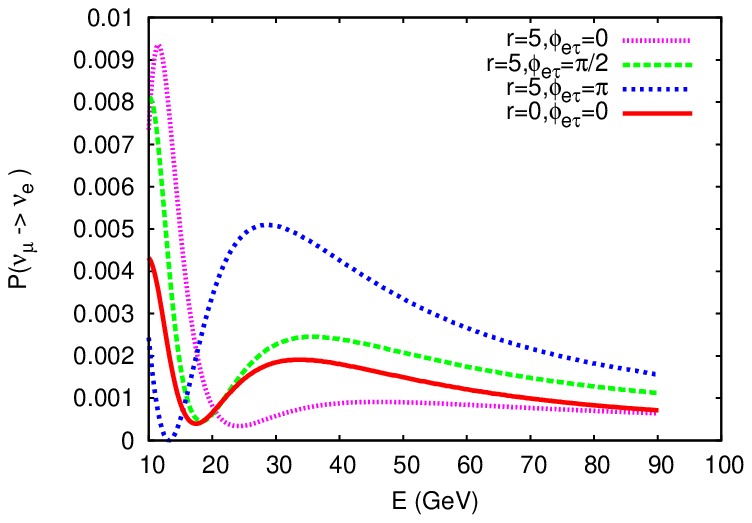}
\end{flushleft}
\begin{flushright}
\vskip -6.5cm
\includegraphics[height=6.cm,width=8cm]{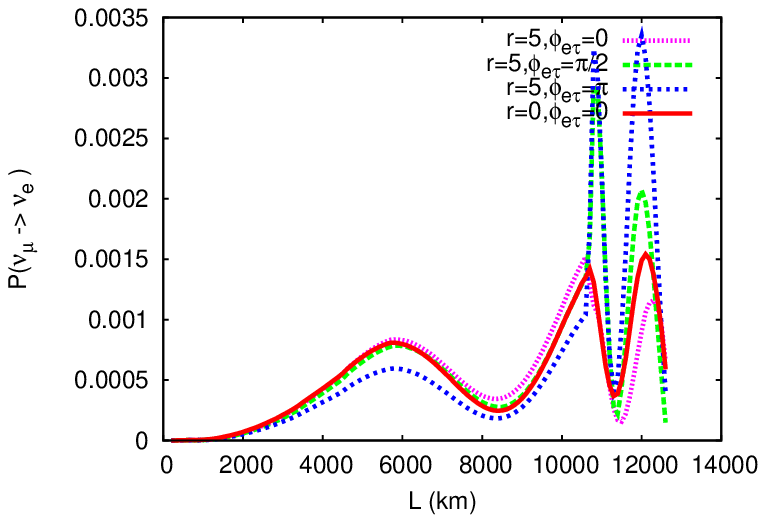}
\end{flushright}
\vskip 0.0cm \caption{\small $P(\nu_\mu \to \nu_e)$ versus
 energy, $L=12000$ km; $P(\nu_\mu \to \nu_e)$ versus
 distance L, $E=50$ GeV. $\phi_{e\mu}=\phi_{\mu \tau}=\pi/2$. Other parameters
 are the same as in Fig. \ref{PVsE}.
 }
 \label{PVsEC}
 \end{figure}

 In Fig. \ref{PVsEB} we see that when $L=12000$km $P_{e\tau}$ is
 reduced when $\varphi_{\tau e}=0$ and is enhanced when $\varphi_{\tau e}=\pi$.
 This phenomenon can be understood by
 considering an interesting case which happens when
 \bea
 \varphi_c+\varphi_1 \approx 2 n \pi, \label{parametric}
 \eea
 where $n$ is an integral.
 Hence $\varphi \approx \varphi_1 + 2 n \pi$.
 This is the region of parametric resonance for oscillation
 with standard matter effect~\cite{param1,param2}. In the presence with non-standard
 matter effect we see that the amplitude is not always enhanced.
 Using Eqs. (\ref{evol2d}) and (\ref{parametric}) we get
 \bea
  A(\nu_e \to \nu_\tau) && \approx \epsilon^0_{\tau e} ~(1-0.098 ~r_{\tau e}
  ~e^{-i \phi_{\tau e}}) (e^{-i \varphi} -1).
 \label{evol2e}
 \eea
 The amplitude is reduced for $\phi_{\tau e}=0$ and is enhanced
 for $\phi_{\tau e}=\pi$. When $r=5$ the transition
 probability is reduced or enhanced by $100\%$.
 When $\phi_{e\tau}=\pi/2$
 $P_{e\tau}$ is not much enhanced. This is understood by noting
 that according to Eq. (\ref{evol2e}) $P_{e\tau}$ is enhanced
 by factor $1+(0.098 r)^2 \approx 1+0.01 r^2$. It is a $25\%$ increase
 when $r=5$. In Fig. \ref{PVsEC} we plot $P(\nu_\mu \to \nu_e)$
 versus energy and the distance L.
 We can also see the effect of $r_{kl}$ and $\phi_{kl}$ in this plot.
 In the right panel of Fig. \ref{PVsEC} significant modifications are
 seen in the second and third peaks.

\section{Conclusions} \label{sec4}

 In summary we have analyzed non-standard matter effect in flavor conversion of
 neutrinos crossing the core of the Earth.
 We have shown that a first order perturbation theory gives a perfect
 description of neutrino oscillation for core-crossing trajectories.
 The analytical description, which uses only zeroth order result,
 gives a good approximation.

 One interesting thing is that there are six physical CP
 violating phases associated with the non-standard matter effect
 when chemical composition changes in matter. This is what happens to
 core crossing neutrinos. It is different from the
 case when chemical composition does not change. In the latter case there
 are only three physical CP violating phases.
 We analyze effect of additional CP violating phases in neutrino oscillation.

 We have shown that due
 to non-standard interaction different chemical composition in the
 core and the mantle ( different $N_n/N_e$ ) can modify neutrino flavor
 conversion by $100\%$.
 We analyze in particular the region of parametric resonance. It is
 shown that in this region the non-standard matter effect does not
 always give enhancement to the amplitude. Depending on the CP violating
 phases the non-standard matter effect reduce or enhance the neutrino flavor
 conversion. The signature of non-standard interactions lies in the
 dependence of the neutrino flavor conversion rate on $E$, the energy of
 neutrinos, and $L$, the length of neutrino trajectory in the Earth.
 To figure out these interactions we need neutrino sources with
 different energies and baselines.

 The analysis presented in the present article shows that core crossing
 neutrino events provide an interesting way to test interactions of
 neutrinos beyond the Standard Model. They also provide an independent
 way to test chemical composition in the Earth. \\

 {\bf Acknowledgment:} The research is supported in part by National
 Science Foundation of China(NSFC), grant 10745003.

\end{document}